\documentclass[11pt,a4paper]{article}%

\usepackage{fancyhdr}    
\usepackage{graphicx}
\usepackage{rotating}

\def\beq{\begin{equation}}
\def\eqn#1{\beq\label{#1}}
\def\eeq{\end{equation}}

\def\bb {\begin {eqnarray}}
\def\eqnn#1{\bb\label{#1}}
\def\ee {\end {eqnarray}}

\def\eqref#1{(\ref{#1})}

\def\ta{{\tilde\alpha}}

\def\hm{{\hat m}}
\def\haa{{\hat a}}
\def\hn{{\hat n}}

\def\nn{\nonumber}
\def\nt{\noindent}

\def\md{\vskip 3mm}

\def\han{{\textstyle{n\over2}}}

\def\llr{\longrightarrow}
\def\({\left(}
\def\){\right)}
\def\eps{\epsilon}
 
\def\lra{\longrightarrow}

\def\ha{{\textstyle{1\over2}}}

  \def\tV{{\tilde V}}

\def\r{\rho}

\def\a{\alpha}
\def\b{\beta}

\def\vr{\vert}

\def\D{{\Delta}}

\def\bbr{{I\!\!R}}
\def\bbn{I\!\!N}
\def\bbc{{C\kern-6.5pt I}\,}
\def\bac{{C\kern-5.5pt I}}
\def\bab{{C\kern-4.5pt I}}

\def\L{\Lambda}
\def\bu{$\bullet$}

\def\riga{-\kern-4pt - \kern-4pt -}
\font\fat=cmsy10 scaled\magstep5

\def\Bbullet{\raise-3pt\hbox{\fat\char"0F}}

\def\ca{{\cal A}}  \def\cc{{\cal C}}
\def\cd{{\cal D}}  \def\cf{{\cal F}}
\def\cg{{\cal G}} \def\ch{{\cal H}} 
 \def\ck{{\cal K}} 
\def\cm{{\cal M}} \def\cn{{\cal N}} 
\def\cp{{\cal P}}  
 \def\ct{{\cal T}}

\def\ido{intertwining differential operator}
\def\idos{intertwining differential operators}

 \def\ha{{\textstyle{\frac{1}{2}}}}
  \def\frh{{\textstyle{\frac{5}{2}}}}
  \def\trh{{\textstyle{\frac{3}{2}}}}

\def\eps{\epsilon}

\def\ca{{\cal A}}
\def\nn{\nonumber}

\def\nt{\noindent}
\def\bu{$\bullet$~~~}
\def\lra{\longleftrightarrow}

\input epsf.tex
\newcount\figno
\def\fig#1#2#3{
\par\begingroup\parindent=0pt\leftskip=1cm\rightskip=1cm\parindent=0pt
\baselineskip=11pt \global\advance\figno by 1 
\epsfxsize=#3 \centerline{\epsfbox{#2}} \vskip 12pt
#1\par
\endgroup\par}
\def\figlabel#1{\xdef#1{\the\figno}}
\def\encadremath#1{\vbox{\hrule\hbox{\vrule\kern8pt\vbox{\kern8pt
\hbox{$\displaystyle #1$}\kern8pt} \kern8pt\vrule}\hrule}}

\begin{document}
 \baselineskip=11pt

\title{Special Reduced Multiplets and Minimal
Representations for Sp(n,R)}
\author{\bf{V.K. Dobrev}\hspace{.25mm}\thanks{\,e-mail address:
dobrev@inrne.bas.bg} \\  \normalsize{Institute for Nuclear Research
and Nuclear Energy,}\\
 \normalsize{Tsarigradsko Chaussee 72, BG-1784 Sofia, Bulgaria}}

\date{}

\maketitle

\begin{abstract}
In the present paper we continue the programme of systematic
construction of invariant differential operators on the example of
the non-compact  groups  Sp(n,R). Earlier in
  arXiv:1205.5521   we gave the main multiplets and the main
 reduced multiplets of indecomposable elementary representations
including the necessary data for all relevant invariant differential
operators.  Here we give the special reduced multiplets and the
minimal representations of Sp(n,R).
\end{abstract}

\section{Introduction}

Invariant differential operators   play very important role in
the description of physical symmetries. For the modern applications of
(super-)differential operators in conformal field theory,
supergravity and string theory we refer, e.g., to \cite{Ter}.

In a recent paper \cite{Dobinv} we started the systematic explicit
construction of invariant differential operators. We gave an
explicit description of the building blocks, namely, the parabolic
subgroups and subalgebras from which the necessary representations
are induced. Thus we have set the stage for study of different
non-compact groups.

 In the present paper we  focus on the groups ~$Sp(n,\bbr)$,
which are very interesting for several reasons. First of all, they
belong to the class of Hermitian symmetric spaces, i.e., the pair
$(G,K)$ is a Hermitian symmetric pair ($K$ is the maximal compact
subgroup of the noncompact semisimple group $G$). These groups have discrete series
representations and highest/lowest weight representations. Further,
~$Sp(n,\bbr)$ ~  belong to a narrower class of groups/algebras,
which we call 'conformal Lie groups or algebras' since they have
very similar properties to the canonical conformal algebras
$so(n,2)$ of $n$-dimensional Minkowski space-time.  This class was
identified from our point of view in \cite{Dobeseven}. Besides
$so(n,2)$ it includes the algebras ~$su(n,n)$, ~$sp(n,\bbr)$,
~$so^*(4n)$, $E_{7(-25)}\,$.  The same class was identified independently
from different considerations and under different names in \cite{FaKo,Guna,Mackder}.

This paper is a   sequel of \cite{Dobspn6}{}, based on Invited
talk at the VII Mathematical Physics Meeting, Belgrade, 9-19.9.2012.
Due to the lack of space we refer to \cite{Dobspn6}
for motivations and extensive list of literature on the subject.

The present paper is organized a follows. In section 2 we give the
preliminaries, actually recalling and adapting facts from
\cite{Dobinv} to the ~$sp(n,\bbr)$~ case. In Section 3 we present
the special reduced multiplets ~$n=2,...,6$ and the \idos\ between
the ERs. In Section 4 we discuss the general features of our results
which generalize for arbitrary $n$. We also present an Outlook.

\section{Preliminaries}

\nt Let ~$n\geq 2$. Let ~$\cg ~=~ sp(n,\bbr)$, the split real form
of ~$sp(n,\bbc)=\cg^\bac$. The maximal compact subgroup of ~$\cg$~
is ~$\ck \cong u(1)\oplus su(n)$.

We choose a ~{\it maximal} parabolic ~$\cp=\cm\ca\cn$~ such that
~$\ca\cong so(1,1)$, while the factor ~$\cm = ~sl(n,\bbr)$~ has the
same finite-dimensional (nonunitary) representations as the
finite-dimensional (unitary) representations of  the semi-simple
subalgebra $su(n)$ of   ~$\ck\,$. Thus, these induced
representations are representations of finite $\ck$-type \cite{HC}.
Note also that ~$\ck^\bac \cong u(1)^\bac \oplus sl(n,\bbc) \cong
\cm^\bac \oplus \ca^\bac$. Finally, note that ~$\dim_\bbr\,\cn =
n(n+1)/2$.

We label   the signature of the ERs of $\cg$   as follows:
\eqn{sgnd}  \chi ~=~ \{\, n_1\,, \ldots,\, n_{n-1}\, ;\, c\, \} \ ,
\qquad n_j \in \bbn\ , \quad c = d- (n+1)/2 \eeq where the last
entry of ~$\chi$~ labels the characters of $\ca\,$, and the first
$n-1$ entries are labels of the finite-dimensional nonunitary irreps
of $\cm\,$, (or of the finite-dimensional unitary irreps of
~$su(n)$).

Below we shall use the following conjugation on the
finite-dimensional entries of the signature: \eqn{conu}
(n_1,\ldots,n_{n-1})^* ~\doteq~ (n_{n-1},\ldots,n_{1}) \eeq

We call the above induced representations ~$\chi =$
Ind$^G_{\cp}(\mu\otimes\nu \otimes 1)$~ ~{\it  elementary
representations} \cite{DMPPT} of $G=Sp(n,\bbr)$. (These are
called {\it generalized principal series representations} (or {\it
limits thereof}) in \cite{Knapp}.)
 Their spaces of functions are:  \begin {eqnarray}\label{func}
\cc_\chi ~&=&~ \{ \cf \in C^\infty(G,V_\mu) ~ \vr ~ \cf (g\hm\haa\hn) ~=~
e^{-\nu(H)} \cdot D^\mu(\hm^{-1})\, \cf (g) \} \nn\end{eqnarray} where
~$\haa= \exp(H)$, ~$H\in\ca\,$, ~$\hm\in M=SL(n,\bbr)$, ~$\hn\in
N=\exp\cn$. The representation action is the {\it left regular
action}:  \begin{equation}\label{lrega} (\ct^\chi(g)\cf) (g') ~=~
\cf (g^{-1}g') ~, \quad g,g'\in G\ .\end{equation}

\bu An important ingredient in our considerations are the ~{\it \it
highest/lowest weight representations}~ of ~$\cg^\bac$. These can be
realized as (factor-modules of) Verma modules ~$V^\L$~ over
~$\cg^\bac$, where ~$\L\in (\ch^\bac)^*$, ~$\ch^\bac$ is a Cartan
subalgebra of ~$\cg^\bac$, weight ~$\L = \L(\chi)$~ is determined
uniquely from $\chi$ \cite{Dob}{}.

Actually, since our ERs are induced from finite-dimensional
representations of ~$\cm$~   the Verma modules are
always reducible. Thus, it is more convenient to use ~{\it
generalized Verma modules} ~$\tV^\L$~ such that the role of the
highest/lowest weight vector $v_0$ is taken by the
(finite-dimensional) space ~$V_\mu\,v_0\,$. For the generalized
Verma modules (GVMs) the reducibility is controlled only by the
value of the conformal weight $d$, or the parameter $c$. Relatedly, for the \idos{} only
the reducibility w.r.t. non-compact roots is essential.

\bu Another  main ingredient of our approach is as follows. We group the
(reducible) ERs with the same Casimirs in sets called ~{\it
multiplets} \cite{Dobmul}{}. The multiplet corresponding to fixed values of the
Casimirs may be depicted as a connected graph, the {\it vertices} of which
correspond to the reducible ERs and the {\it lines (arrows)}  between the vertices
correspond to intertwining operators.
The multiplets contain explicitly all the data necessary to
construct the \idos{}. Actually, the data for each \ido{} consists
of the pair ~$(\b,m)$, where $\b$ is a (non-compact) positive root
of ~$\cg^\bac$, ~$m\in\bbn$, such that the {\it BGG  Verma module
reducibility condition} \cite{BGG} (for highest weight modules) is fulfilled:
\begin{equation}\label{bggr} (\L+\r, \b^\vee ) ~=~ m \ , \quad \b^\vee \equiv 2 \b
/(\b,\b) \ \end{equation} where $\r$ is half the sum of the positive roots of
~$\cg^\bac$. When the above holds then the Verma module with shifted
weight ~$V^{\L-m\b}$ (or ~$\tV^{\L-m\b}$ ~ for GVM and $\b$
non-compact) is embedded in the Verma module ~$V^{\L}$ (or
~$\tV^{\L}$). This embedding is realized by a singular vector
~$v_s$~  expressed by a polynomial ~$\cp_{m,\b}(\cg^-)$~ in the universal
enveloping algebra ~$(U(\cg_-))\ v_0\,$, ~$\cg^-$~ is the subalgebra
of ~$\cg^\bac$ generated by the negative root generators \cite{Dix}{}.
 More explicitly, \cite{Dob}, ~$v^s_{m,\b} = \cp_{m,\b}\, v_0$ (or ~$v^s_{m,\b} =
 \cp_{m,\b}\, V_\mu\,v_0$ for GVMs).\\
   Then there exists \cite{Dob} an ~{\it \ido{}}~ of order ~$m=m_\b$~:
\begin{equation}\label{invop}
  \cd_{m,\b} ~:~ \cc_{\chi(\L)}
~\llr ~ \cc_{\chi(\L-m\b)} \end{equation} given explicitly by: \begin{equation}\label{singvv}
 \cd_{m,\b} ~=~ \cp_{m,\b}(\widehat{\cg^-})  \end{equation} where
~$\widehat{\cg^-}$~ denotes the {\it right action} on the functions
~$\cf\,$.

Thus, in each such situation we have an ~{\it invariant differential equation}~ of order ~$m=m_\b$~:
\begin{equation}\label{invde} \cd_{m,\b}\ f ~=~ f' \ , \qquad f \in \cc_{\chi(\L)} \ , \quad
f' \in \cc_{\chi(\L-m\b)} \ .\end{equation}

In most of these situations the invariant operator ~$\cd_{m,\b}$~ has a non-trivial invariant
kernel in which a subrepresentation of $\cg$ is realized. Thus, studying the equations
with trivial RHS:
\begin{equation}\label{invdec} \cd_{m,\b}\ f ~=~ 0 \ , \qquad f \in \cc_{\chi(\L)} \ ,
   \end{equation}
is also very important. For example, in many physical applications
 in the case of first order differential operators,
i.e., for ~$m=m_\b = 1$, equations (\ref{invdec})
are called ~{\it conservation laws}, and the elements ~$f\in \ker \cd_{m,\b}$~
are called ~{\it conserved currents}.  

The ERs in the multiplet are related also by intertwining integral
  operators. The  integral operators were introduced by
Knapp and Stein \cite{KnSt}. In fact, these operators are defined
for any ER, not only for the reducible ones, the general action
being: \eqnn{ksop} & G_{KS} ~:~ \cc_\chi ~ \llr ~ \cc_{\chi'} \ ,\cr
&\chi ~=~ \{\, n_1,\ldots,n_{n-1} \,;\, c\, \} \ , \qquad \chi' ~=~
\{\, (n_1,\ldots,n_{n-1})^* \,;\, -c\, \}  \ee
The above action on the signatures is also called restricted Weyl reflection, since it represents
the nontrivial element of the 2-element restricted Weyl group which arises canonically
with every maximal parabolic subalgebra.
Generically, the Knapp-Stein operators can be normalized
so that indeed ~$G_{KS} \circ G_{KS} = {\rm Id}_{\cc_\chi}\,$. However, this usually fails exactly for
the reducible ERs that form the multiplets, cf., e.g., \cite{DMPPT}.

Further, we need more explicitly the root system of the algebra
~$sp(n,F)$, $F=\bbc,\bbr$.
In terms of the orthonormal basis $\eps_i\,$, ~$i=1,\ldots,n$, the
 positive roots are given by \eqn{spnrpos} \D^+ = \{ \eps_i \pm \eps_j, ~1 \leq
i <j \leq n; ~2\eps_i, 1 \leq i \leq n\} ,  \eeq while the simple
roots are: \eqn{spnrsmp} \pi = \{\a_i = \eps_i - \eps_{i+1}, ~1 \leq
i \leq n - 1; \ \a_n = 2\eps_n\} \eeq With our choice of
normalization of
  the long roots  ~$2\eps_k$~ have
length 4, while the short roots ~$\eps_i \pm \eps_j$~ have length 2.

From these the compact roots are those that form (by restriction)
the root system of the semisimple part of ~$\ck^\bac$, the rest are
noncompact, i.e., \eqnn{spnrcnc}  {\rm compact:}&~~~ \a_{ij}
~\equiv~\eps_i  -\eps_j \ , \quad 1 \leq i < j \leq n \ ,
  \cr {\rm noncompact:}&~~~ \b_{ij} ~\equiv~ \eps_i
+\eps_j \ ,  ~~ 1 \leq i \leq j \leq n \  \ee Thus, the only
non-compact simple root is ~$\a_n=\b_{nn}\,$.

Further, we shall use the so-called Dynkin labels: \eqn{dynk} m_i
~\equiv~ (\L+\r,\a^\vee_i)  \ , \quad i=1,\ldots,n,\eeq where ~$\L =
\L(\chi)$, ~$\r$ is half the sum of the positive roots of
~$\cg^\bac$.

We shall use also   the so-called Harish-Chandra parameters:
\eqn{dynhc} m_\b \equiv (\L+\r, \b )\ ,\eeq where $\b$ is any
positive root of $\cg^\bac$. These parameters are redundant, since
they are expressed in terms of the Dynkin labels, however,   some
statements are best formulated in their terms. In particular, in the
case of the noncompact roots we have: \eqn{hclab} m_{\b_{ij}} ~=~
\Big( \sum_{s=i}^n + \sum_{s=j}^n \Big)   m_s \ , ~~~ i<j \ ; \qquad
m_{\b_{ii}} ~=~  \sum_{s=i}^n    m_s  \eeq

Finally,  we  give the correspondence between the signatures $\chi$
and the highest weight $\L$. The explicit connection is: \eqn{rela}
n_i = m_i \ , \quad  c ~=~ -\ha (m_\ta + m_n) ~=~ -\,\ha( m_1+\cdots
+ m_{n-1} + 2m_n  )\eeq where ~$\ta ~=~ \b_{11}$~ is the highest
root.

\section{Special reduced multiplets and minimal UIRs}

There are several types of multiplets: the main type, (which
contains maximal number of ERs/GVMs, the finite-dimensional and the
discrete series representations), and various  reduced types of
multiplets.  The multiplets of the main type are in 1-to-1 correspondence
with the finite-dimensional irreps of ~$sp(n,\bbr)$, i.e., they will
be labelled by  the ~$n$~ positive Dynkin labels    ~$m_i\in\bbn$.
As we mentioned, each main multiplet contains ~$2^n$ ERs/GVMs.  It
is difficult to give explicitly the multiplets for general ~$n$.
Thus, in the paper \cite{Dobspn6} we gave  for $sp(n,\bbr)$, $n=6$, the
 the main type of multiplets and the main
reduced types (which depend on $n-1$ parameters).
In fact, this gives by reduction also the cases for
~$n<6$, since   the main multiplet for fixed ~$n$~ coincides with
one reduced case for ~$n+1$.

In the present paper we give for ~$n=2,...,6$ the special reduced multiplets which depend
on ~$n-1$ positive integers and one positive odd integer.

\subsection{The case sp(2,$\bbr$)}

The material of this subsection is contained in \cite{Dobads} and \cite{Dobpeds} and is given here
to set the stage for the higher rank cases.

The main multiplets $R^2_m$ of $sp(2,\bbr)$ contain
$4 (=2^2)$ ERs/GVMs whose signatures were given in \cite{Dobads}
in the following pair-wise manner:
\eqnn{tabltwo}
&\chi_0^\pm ~=~ \{\,
 m_1\,;\,\trh\pm\ha (m_1+2m_{2}) \,\}  \\
  &\chi_a^\pm ~=~ \{\,
  m_{1}+2m_2\,;\,\trh\pm\ha m_{1}\,\} \nn\ee
The multiplets are given explicitly in Fig. 1.  where we use the
notation: ~$\L^\pm = \L(\chi^\pm)$.  Each \ido\ is represented by an
arrow accompanied by a symbol ~$i_{j...k}$~ encoding the root
~$\b_{j...k}$~ and the number $m_{\b_{j...k}}$ which is involved in
the BGG criterion. This notation is used to save space, but it can
be used due to
 the fact that only \idos\ which are
non-composite are displayed, and that the data ~$\b,m_\b\,$, which
is involved in the embedding ~$V^\L \lra V^{\L-m_\b,\b}$~ turns out
to involve only the ~$m_i$~ corresponding to simple roots, i.e., for
each $\b,m_\b$ there exists ~$i = i(\b,m_\b,\L)\in \{
1,\ldots,2n-1\}$, such that ~$m_\b=m_i\,$. Hence the data
~$\b_{j...k}\,$,~$m_{\b_{j...k}}$~ is represented by ~$i_{j...k}$~
on the arrows.

The pairs ~$\L^\pm$~ of \eqref{tablsps}
in Figure 1. are symmetric   w.r.t. to the bullet in the
middle of the figure - this represents the Weyl symmetry realized by
two Knapp-Stein integral operators \eqref{ksop}:
\eqn{ksop2}  G^\pm_{KS} ~:~ \cc_{\L^\mp}  ~ \llr ~ \cc_{\L^\pm} \ . \eeq

In \cite{Dobads} and \cite{Dobpeds} the same multiplet was given with the Knapp-Stein operators
displayed explicitly - see Figure 2. The differential operators are denoted by arrows, the integral operators
 - by   dashed arrows.  We see that the Knapp-Stein operator from ~$\cc_{\L^-_a}$~ to ~$\cc_{\L^+_a}$~ is
 degenerated to the differential operator denoted by ~$1_{12}\,$. Certainly, the latter degeneration is seen
 already in Figure 1 from the symmetry w.r.t. the bullet in the centre of the figure.

For further use we denote  as ~$\cd^\pm$~ the invariant subspaces of ~$\cc_{\L^\pm}$~  such that
 ~$\cd^\pm$~ is the image of ~$G^\pm_{KS}$~ and
the kernel of ~$G^\mp_{KS}\,$.  This feature is common for all $sp(n)$.

The special reduced multiplets $R^2_s$ also contain $4$  ERs/GVMs whose
signatures are given in the following pair-wise manner:
\eqnn{tabltwos}
&\chi_0^\pm ~=~ \{\,
 m_1\,;\,\trh\pm\ha (m_1+\mu) \,\}  \\
  &\chi_a^\pm ~=~ \{\,
  m_{1}+\mu\,;\,\trh\pm\ha m_{1}\,\}  \nn\ee
where ~$\mu\in 2\bbn-1$.
The multiplets are given explicitly in Fig. 3.

Furthermore, as in the main multiplet  these
representations have all the same Casimirs, but
none of them contains a finite-dimensional irrep. Neither the are
related   as in the quartet since there are no analogs of the
operators from $\chi^-_{0}$ to $\chi^-_{a}$ or from $\chi^+_{a}$ to
$\chi^+_{0}$.  In   terms of the quartet diagram on Figure 2, only the
horizontal lines/arrows remain valid.

Thus, although superficially there are three connected components in the Figure, taking
 into account the Knapp-Stein operators each multiplet contains two connected components, or submultiplets:\\
\bu a doublet consisting of ~$\L^\pm_0\,$;\\
\bu a doublet consisting of ~$\L^\pm_a\,$.

  Next we recall from \cite{GeV} following the exposition in \cite{Dobpeds} that   each ER ~$\chi^+_0$~ of \eqref{tabltwo}
contains both a holomorphic discrete series irrep  and its conjugate anti-holomorphic discrete series irrep.
The direct sum  of the   representation spaces of these two irreps is the invariant subspace ~$\cd^+_0$~ of ~$\cc_{\L^+_0}$~
mentioned above.
The statement about the disposition of the (anti)holomorphic discrete series in the the ERs ~$\chi^+_0$~ of the main multiplets
is valid for all ~$n$. In particular, the corresponding lowest value of the conformal weight is ~$d=n+1$~
which equals the dimension of the corresponding space-time.

For the lack of space we do not discuss the disposition of the   (anti)holo\-morphic discrete series, the
 first reduction point (FRP), and other positive energy irreps except the minimal UIRs \cite{Koba}.
 Here the latter  are two special irreps discovered by Dirac \cite{Dirac}
and called 'singletons'  or 'Di' and 'Rac' in  \cite{Fro}:
\eqn{unising}
 {\rm Rac} ~:~ (d,s_0) ~=~ (1/2,0) ~,  \qquad  {\rm Di} ~:~
(d,s_0) ~=~ (1,1/2) ~. \eeq
Both are situated in the special reduced multiplets:
the 'Rac' is situated in the ER/GVM ~$\chi^-_0$~ of \eqref{tabltwos}  with ~$m_1=\mu=1$, while the 'Di'
 in the ER/GVM ~$\chi^-_a$~ of \eqref{tabltwos}  with ~$m_1=\mu=1$.

\md

\subsection{The case sp(3,$\bbr$)}

\nt  The main multiplets $R^3_m$ of $sp(3,\bbr)$ contain
$8 (=2^3)$ ERs/GVMs whose signatures are given in the following pair-wise manner:
\eqnn{tabltr}
&\chi_0^\pm ~=~ \{\, (
 m_1,
 m_2)^\pm\,;\,2\pm\ha (m_{12}+2m_3) \,\}  \\
 &\chi_a^\pm ~=~ \{\, (
 m_1,m_2+2m_{3})^\pm\,;\,2\pm\ha m_{12}\,\} \cr
&\chi_b^\pm ~=~ \{\, (
  m_{12},
 m_{2}+2m_3)^\pm\,;\,2\pm\ha m_{1}\,\} \cr
 &\chi_c^\pm ~=~ \{\, (
 m_{2},
 m_{12}+2 m_{3})^\pm\,;\,2\mp\ha m_{1}\,\} \nn\ee
 and the notation ~$(...)^\pm$~ employs the   conjugation
(\ref{conu})~:  $$
  (n_1,...,n_{n-1})^- ~=~ (n_1,...,n_{n-1})\ , \qquad
(n_1,...,n_{n-1})^+ ~=~  (n_1,...,n_{n-1})^*  $$
The multiplets are given explicitly in Fig. 4.

The special reduced multiplets $R^3_s$ also contain $8$  ERs/GVMs whose
signatures are given in the following pair-wise manner:
\eqnn{tabltrs}
&\chi_0^\pm ~=~ \{\, (
 m_1,
 m_2)^\pm\,;\,2\pm\ha (m_{12}+\mu) \,\}  \\
 &\chi_a^\pm ~=~ \{\, (
 m_1,m_2+\mu)^\pm\,;\,2\pm\ha m_{12}\,\} \cr
&\chi_b^\pm ~=~ \{\, (
  m_{12},
 m_{2}+\mu)^\pm\,;\,2\pm\ha m_{1}\,\} \cr
 &\chi_c^\pm ~=~ \{\, (
 m_{2},
 m_{12}+\mu)^\pm\,;\,2\mp\ha m_{1}\,\}  \nn\ee
where ~$\mu\in 2\bbn-1$.
The multiplets are given explicitly in Fig. 5.

 Taking  into account the Knapp-Stein operators then each such multiplet contains two
 connected components, or submultiplets:\\
\bu a doublet consisting of ~$\L^\pm_0\,$;\\
\bu a submultiplet of 6 ERs/GVMs consisting of ~$\L^\pm_a\,$, ~$\L^\pm_b\,$, ~$\L^\pm_c\,$.

There are three minimal UIRs situated in ~$\L^-_0$, ~$\L^-_a$ and ~$\L^+_c\,$  ~with $m_1=m_2=\mu=1$~:\\
\bu The one in ~$\L^-_0$~ has   trivial $su(3)$ irrep and ~$d=\ha\,$.\\
\bu The one in ~$\L^-_a$~ has  three-dimensional $su(3)$ irrep and ~$d=1$.\\
\bu The one in ~$\L^+_c$~ has  six-dimensional $su(3)$ irrep and ~$d=\trh$.\\
Note that ~$\L^-_a$ and $\L^+_c$~ are in the same connected component - this feature is commented in Section 4.

\subsection{The case sp(4,$\bbr$)}

\nt  The main multiplets $R^4_m$ of $sp(4,\bbr)$ contain
$16 (=2^4)$ ERs/GVMs whose signatures are given in the following pair-wise manner:
\eqnn{tablfr}
&\chi_0^\pm ~=~ \{\, ( 
 m_1,
 m_2,
 m_3)^\pm\,;\,\pm\ha(m_{13} + 2m_4  )\,\}  \\
&\chi_a^\pm ~=~ \{\, ( 
 m_1,
 m_2,
 m_3+2m_4)^\pm\,;\,\pm\ha m_{13}\,\} \cr
&\chi_b^\pm ~=~ \{\, ( 
 m_1,
 m_{23},
 m_3+2m_4)^\pm\,;\,\pm\ha m_{12}\,\} \cr
&\chi_c^\pm ~=~ \{\, (  
  m_{12},
 m_{3},
 m_{23}+2m_4)^\pm\,;\,\pm\ha m_{1}\,\} \cr
&\chi_d^\pm ~=~ \{\, (  
 m_{2},
 m_{3},
 m_{13}+2m_4)^\pm\,;\,\mp\ha m_{1}\,\} \cr
 &\chi_e^\pm ~=~ \{\, (  
 m_{2},
m_{3}+2m_4,
 m_{13}
 )^\pm\,;\,\mp\ha m_{1}\,\}\cr
&\chi_{f}^\pm ~=~ \{\, (  
 m_{12},
 m_{3}+2m_4,
 m_{23})^\pm\,;\,\pm\ha m_{1}\,\} \cr
&\chi_{g}^\pm ~=~ \{\, ( 
 m_1,
 m_{23}+2m_4,
 m_3)^\pm\,;\,\pm\ha m_{12}\,\}
  \nn\ee
The multiplets are given explicitly in Fig. 3.

The special reduced multiplets $R^4_s$ also contain $16$  ERs/GVMs whose
signatures are given in the following pair-wise manner:
\eqnn{tablfrs}
&\chi_0^\pm ~=~ \{\, ( 
 m_1,
 m_2,
 m_3)^\pm\,;\,\frh\pm\ha(m_{13} + \mu  )\,\}  \\
&\chi_a^\pm ~=~ \{\, ( 
 m_1,
 m_2,
 m_3+\mu)^\pm\,;\,\frh\pm\ha m_{13}\,\} \cr
&\chi_b^\pm ~=~ \{\, ( 
 m_1,
 m_{23},
 m_3+\mu)^\pm\,;\,\frh\pm\ha m_{12}\,\} \cr
&\chi_c^\pm ~=~ \{\, (  
  m_{12},
 m_{3},
 m_{23}+\mu)^\pm\,;\,\frh\pm\ha m_{1}\,\} \cr
&\chi_d^\pm ~=~ \{\, (  
 m_{2},
 m_{3},
 m_{13}+\mu)^\pm\,;\,\frh\mp\ha m_{1}\,\} \cr
&\chi_e^\pm ~=~ \{\, (  
 m_{2},
m_{3}+\mu,
 m_{13}
 )^\pm\,;\,\frh\mp\ha m_{1}\,\} \cr
&\chi_{f}^\pm ~=~ \{\, (  
 m_{12},
 m_{3}+\mu,
 m_{23})^\pm\,;\,\frh\pm\ha m_{1}\,\} \cr
&\chi_{g}^\pm ~=~ \{\, ( 
 m_1,
 m_{23}+\mu,
 m_3)^\pm\,;\,\frh\pm\ha m_{12}\,\}
   \nn\ee
where ~$\mu\in 2\bbn-1$.
The multiplets are given explicitly in Fig. 3s.

 Taking  into account the Knapp-Stein operators then each such multiplet contains three
 connected components, or submultiplets:\\
\bu a doublet consisting of ~$\L^\pm_0\,$;\\
\bu a submultiplet of 8 ERs/GVMs  consisting of ~$\L^\pm_a\,$, ~$\L^\pm_b\,$, ~$\L^\pm_c\,$, ~$\L^\pm_d\,$;\\
\bu a submultiplet of 6 ERs/GVMs consisting of   ~$\L^\pm_{e}\,$, ~$\L^\pm_{f}\,$, ~$\L^\pm_{g}\,$.

There are four minimal UIRs situated in ~$\L^-_0$, ~$\L^-_a$, ~$\L^+_d$, ~ and ~$\L^-_g$ ~with $m_1=m_2=m_3=\mu=1$~:\\
\bu The one in ~$\L^-_0$~ has  have trivial $su(4)$ irrep and ~$d=\ha\,$.\\
\bu The one in ~$\L^-_a$~ has  fundamental $su(4)$ irrep and ~$d=1$.\\
\bu The one in ~$\L^-_g$~ has  20-dimensional $su(4)$ irrep and ~$d=\trh$.\\
\bu The one in ~$\L^+_d$~ has  another 20-dimensional $su(4)$ irrep and ~$d=2$.\\
Note that ~$\L^-_a$ and $\L^+_d$~ are in the same connected component.

\subsection{The case sp(5,$\bbr$)}

\nt  The main multiplets $R^5_m$ of $sp(5,\bbr)$ contain
$32 (=2^5)$ ERs/GVMs whose signatures were given in \cite{Dobspn6}.
The special reduced multiplets $R^5_s$ also contain $32$  ERs/GVMs
whose signatures are given in the following pair-wise manner:
\eqnn{tablfv}
&\chi_0^\pm ~=~ \{\, ( 
 m_1,
 m_2,
 m_3,
 m_4)^\pm\,;\,\pm\ha(m_{14} + \mu  )\,\} \\
&\chi_a^\pm ~=~ \{\, ( 
 m_1,
 m_2,
 m_3,
 m_4+\mu)^\pm\,;\,\pm\ha m_{14}\,\} \cr
&\chi_b^\pm ~=~ \{\, ( 
 m_1,
 m_2,
 m_{34},
 m_4+\mu)^\pm\,;\,\pm\ha m_{13}\,\} \cr
&\chi_c^\pm ~=~ \{\, ( 
 m_1,
 m_{23},
 m_{4},
 m_{34}+\mu)^\pm\,;\,\pm\ha m_{12}\,\} \cr
&\chi_{c'}^\pm ~=~ \{\, ( 
 m_1,
 m_2,
 m_{34}+\mu,
 m_4)^\pm\,;\,\pm\ha m_{13}\,\} \cr
&\chi_d^\pm ~=~ \{\, ( 
 m_{12},
 m_{3},
 m_{4},
 m_{24}+\mu)^\pm\,;\,\pm\ha m_{1}\,\} \cr
 &\chi_{d'}^\pm ~=~ \{\, ( 
 m_1,
 m_{23},
 m_{4}+\mu,
 m_{34})^\pm\,;\,\pm\ha m_{12}\,\} \cr
 &\chi_e^\pm ~=~ \{\, ( 
 m_{2},
 m_{3},
 m_{4},
 m_{14}+\mu)^\pm\,;\,\mp\ha m_{1}\,\} \cr
&\chi_{e'}^\pm ~=~ \{\, ( 
 m_{12},
 m_{3},
 m_{4}+\mu,
 m_{24})^\pm\,;\,\pm\ha m_{1}\,\} \cr
 &\chi_{e''}^\pm ~=~ \{\, ( 
 m_1,
 m_{24},
 m_{4}+\mu,
 m_{3})^\pm\,;\,\pm\ha m_{12}\,\} \cr
&\chi_f^\pm ~=~ \{\, (  
 m_{2},
 m_{3},
 m_{4}+\mu,
 m_{14}
 )^\pm\,;\,\mp\ha m_{1}\,\} \cr
&\chi_{f'}^\pm ~=~ \{\, ( 
 m_{12},
 m_{34},
 m_{4}+\mu,
 m_{23})^\pm\,;\,\pm\ha m_{1}\,\} \cr
&\chi_{f''}^\pm ~=~ \{\, ( 
 m_{1},
m_{24}+\mu,
 m_{4},
 m_{3})^\pm\,;\,\pm\ha m_{12}\,\} \cr
&\chi_g^\pm ~=~ \{\, ( 
 m_{2},
 m_{34},
 m_{4}+\mu,
 m_{13})^\pm\,;\,\mp\ha m_{1}\,\} \cr
&\chi_{g'}^\pm ~=~ \{\, (  
 m_{12},
m_{34}+\mu,
 m_{4},
 m_{23})^\pm\,;\,\pm\ha m_{1}\,\} \cr
 &\chi_h^\pm ~=~ \{\, ( 
 m_{2},
 m_{34}+\mu,
 m_{4},
 m_{13})^\pm\,;\,\mp\ha m_{1}\,\} \nn\ee
where ~$\mu\in 2\bbn-1$.
The multiplets are given explicitly in Fig. 2s.

 Taking  into account the Knapp-Stein operators then each such multiplet contains three
 connected components, or submultiplets:\\
\bu a doublet consisting of ~$\L^\pm_0\,$;\\
\bu a submultiplet of 10 ERs/GVMs starting with  ~$\L^-_a\,$;\\
\bu a submultiplet of 20 ERs/GVMs starting with  ~$\L^-_{c'}\,$.

 There are five minimal UIRs situated in ~$\L^-_0$, ~$\L^-_a$, ~$\L^+_e$, ~$\L^-_{c'}$ and $\L^-_{f''}$
  ~with $m_1=m_2=m_3=m_4=\mu=1$~:\\
\bu The one in ~$\L^-_0$~ has  have trivial $su(5)$ irrep and ~$d=\ha\,$.\\
\bu The one in ~$\L^-_a$~ has  fundamental $su(5)$ irrep and ~$d=1$.\\
\bu The one in ~$\L^-_{c'}$~ has  50-dimensional $su(5)$ irrep and ~$d=\trh$.\\
\bu The one in ~$\L^-_{f''}$~ has  175-dimensional $su(5)$ irrep and ~$d=2$.\\
\bu The one in ~$\L^+_{e}$~ has  70-dimensional $su(5)$ irrep and ~$d=\frh$.\\
Note that ~$\L^-_a$ and $\L^+_e$~ are in the same connected component, same for
~$\L^-_{c'}$ and $\L^-_{f''}\,$.

\subsection{The case sp(6,$\bbr$)}

The main multiplets $R^6_m$ of $sp(6,\bbr)$ contain
$64 (=2^6)$ ERs/GVMs whose signatures were given in \cite{Dobspn6}
and we omit here for the lack of space.
The special reduced multiplets $R^6_s$ also contain $64$  ERs/GVMs
whose signatures are given in the following pair-wise manner:
\eqnn{tablsps}
&\chi_0^\pm ~=~ \{\, (
 m_1,
 m_2,
 m_3,
 m_4,
 m_5)^\pm\,;\,\pm\ha(m_{15} + \mu  )\,\}  \\
&\chi_a^\pm ~=~ \{\, ( 
 m_1,
 m_2,
 m_3,
 m_4,
 m_5+\mu)^\pm\,;\,\pm\ha m_{15}\,\} \cr
&\chi_b^\pm ~=~ \{\, (  
 m_1,
 m_2,
 m_3,
 m_{45},
 m_5+\mu)^\pm\,;\,\pm\ha m_{14}\,\} \cr
&\chi_c^\pm ~=~ \{\, (
 m_1,
 m_2,
 m_{34},
 m_{5},
 m_{45}+\mu)^\pm\,;\,\pm\ha m_{13}\,\} \cr
&\chi_{c'}^\pm ~=~ \{\, ( 
 m_1,
 m_2,
 m_3,
 m_{45}+\mu,
 m_5)^\pm\,;\,\pm\ha m_{14}\,\} \cr
&\chi_d^\pm ~=~ \{\, ( 
 m_1,
 m_{23},
 m_{4},
 m_{5},
 m_{35}+\mu)^\pm\,;\,\pm\ha m_{12}\,\} \cr
 &\chi_{d'}^\pm ~=~ \{\, (
 m_1,
 m_2,
 m_{34},
 m_{5}+\mu,
 m_{45})^\pm\,;\,\pm\ha m_{13}\,\} \cr
 &\chi_e^\pm ~=~ \{\, ( 
 m_{12},
 m_{3},
 m_{4},
 m_{5},
 m_{25}+\mu)^\pm\,;\,\pm\ha m_{1}\,\} \cr
&\chi_{e'}^\pm ~=~ \{\, (
 m_1,
 m_{23},
 m_{4},
 m_{5}+\mu,
 m_{35})^\pm\,;\,\pm\ha m_{12}\,\} \cr
 &\chi_{e''}^\pm ~=~ \{\, (
 m_1,
 m_2,
 m_{35},
 m_{5}+\mu,
 m_{4})^\pm\,;\,\pm\ha m_{13}\,\} \cr
&\chi_f^\pm ~=~ \{\, (
 m_{2},
 m_{3},
 m_{4},
 m_{5},
 m_{15}+\mu)^\pm\,;\,\mp\ha m_{1}\,\} \cr
&\chi_{f'}^\pm ~=~ \{\, ( 
 m_{12},
 m_{3},
 m_{4},
 m_{5}+\mu,
 m_{25})^\pm\,;\,\pm\ha m_{1}\,\} \cr
&\chi_{f''}^\pm ~=~ \{\, (
 m_{1},
 m_{23},
 m_{45},
 m_{5}+\mu,
 m_{34})^\pm\,;\,\pm\ha m_{12}\,\} \cr
&\chi_{f'''}^\pm ~=~ \{\, (
 m_1,
 m_2,
 m_{35}+\mu,
 m_{5},
 m_{4})^\pm\,;\,\pm\ha m_{13}\,\} \cr
&\chi_g^\pm ~=~ \{\, ( 
 m_{2},
 m_{3},
 m_{4},
 m_{5}+\mu,
 m_{15})^\pm\,;\,\mp\ha m_{1}\,\} \cr
&\chi_{g'}^\pm ~=~ \{\, ( 
 m_{12},
 m_{3},
 m_{45},
 m_{5}+\mu,
 m_{24})^\pm\,;\,\pm\ha m_{1}\,\} \cr
&\chi_{g''}^\pm ~=~ \{\, (
 m_1,
 m_{23},
 m_{45}+\mu,
 m_{5},
 m_{34})^\pm\,;\,\pm\ha m_{12}\,\} \cr
 &\chi_h^\pm ~=~ \{\, ( 
 m_{2},
 m_{3},
 m_{45},
 m_{5}+\mu,
 m_{14})^\pm\,;\,\mp\ha m_{1}\,\} \cr
 &\chi_{h'}^\pm ~=~ \{\, ( 
 m_{12},
 m_{3},
 m_{45}+\mu,
 m_{5},
 m_{24})^\pm\,;\,\pm\ha m_{1}\,\} \cr
&\chi_{h''}^\pm ~=~ \{\, (
 m_2,
 m_{3},
 m_{45}+\mu,
 m_{5},
 m_{14})^\pm\,;\,\mp\ha m_{1}\,\} \cr
 &\chi_j^\pm ~=~ \{\, ( 
 m_{2},
 m_{34},
 m_{5},
 m_{45}+\mu,
 m_{13})^\pm\,;\,\mp\ha m_{1}\,\} \cr
 &\chi_{j'}^\pm ~=~ \{\, (
 m_{12},
 m_{34},
 m_{5},
 m_{45}+\mu,
 m_{23})^\pm\,;\,\pm\ha m_{1}\,\} \cr
 &\chi_{j''}^\pm ~=~ \{\, (
 m_{1},
 m_{24},
 m_{5},
 m_{45}+\mu,
 m_{3})^\pm\,;\,\pm\ha m_{12}\,\} \cr
 &\chi_k^\pm ~=~ \{\, ( 
 m_{2},
 m_{34},
 m_{5}+\mu,
 m_{45},
 m_{13})^\pm\,;\,\mp\ha m_{1}\,\} \cr
 &\chi_{k'}^\pm ~=~ \{\, ( 
 m_{12},
 m_{34},
 m_{5}+\mu,
 m_{45},
 m_{23})^\pm\,;\,\pm\ha m_{1}\,\} \cr
 &\chi_{k''}^\pm ~=~ \{\, (
 m_{1},
 m_{24},
 m_{5}+\mu,
 m_{45},
 m_{3})^\pm\,;\,\pm\ha m_{12}\,\} \cr
 &\chi_\ell^\pm ~=~ \{\, ( 
 m_{2},
 m_{35},
 m_{5}+\mu,
 m_{4},
 m_{13})^\pm\,;\,\mp\ha m_{1}\,\} \cr
 &\chi_{\ell'}^\pm ~=~ \{\, ( 
 m_{12},
 m_{35},
 m_{5}+\mu,
 m_{4},
 m_{23})^\pm\,;\,\pm\ha m_{1}\,\} \cr
 &\chi_{\ell''}^\pm ~=~ \{\, (
 m_{1},
 m_{25},
 m_{5}+\mu,
 m_{4},
 m_{3})^\pm\,;\,\pm\ha m_{12}\,\} \cr
&\chi_m^\pm ~=~ \{\, ( 
 m_{2},
 m_{35}+\mu,
 m_{5},
 m_{4},
 m_{13})^\pm\,;\,\mp\ha m_{1}\,\} \cr
 &\chi_{m'}^\pm ~=~ \{\, ( 
 m_{12},
 m_{35}+\mu,
 m_{5},
 m_{4},
 m_{23})^\pm\,;\,\pm\ha m_{1}\,\} \cr
 &\chi_{m''}^\pm ~=~ \{\, (
 m_{1},
 m_{25}+\mu,
 m_{5},
 m_{4},
 m_{3})^\pm\,;\,\pm\ha m_{12}\,\} \nn\ee where ~$\mu\in 2\bbn-1$.

Taking into account the Knapp-Stein operators then each such multiplet contains  four
 connected components, or submultiplets:\\
\bu a doublet consisting of ~$\L^\pm_0\,$;\\
\bu a submultiplet of 12 ERs/GVMs starting with  ~$\L^-_a\,$;\\
\bu a submultiplet of 30 ERs/GVMs starting with  ~$\L^-_{c'}\,$;\\
\bu a submultiplet of 20 ERs/GVMs starting with ~$\L^-_{f'''}\,$.

There are six minimal UIRs with $m_1=m_2=m_3=m_4=m_5=\mu=1$~:\\
\bu The one in ~$\L^-_0$~ has  have trivial $su(6)$ irrep and ~$d=\ha\,$.\\
\bu The one in ~$\L^-_a$~ has  fundamental $su(6)$ irrep and ~$d=1$.\\
\bu The one in ~$\L^-_{c'}$~ has 105-dimensional $su(6)$ irrep and ~$d=\trh$.\\
\bu The one in ~$\L^-_{f'''}$~ has 980-dimensional $su(6)$ irrep and ~$d=2$.\\
\bu The one in ~$\L^-_{m''}$~ has 1764-dimensional $su(6)$ irrep and ~$d=\frh$.\\
\bu The one in ~$\L^+_{f}$~ has 252-dimensional $su(6)$ irrep and ~$d=3$.\\
Note that ~$\L^-_a$ and $\L^+_f$~ are in the same connected component, same for
~$\L^-_{c'}$ and $\L^-_{m''}\,$.

\section{Summary, discussion of results and outlook}

Each special reduced multiplet of ~$sp(n,\bbr)$~ contains ~$[\han]+1$~ connected components, or submultiplets, although
if we do not take into account the Knapp-Stein operators then there would be  ~$n+1$~ connected components.

All the ERs/GVMs in a special reduced multiplet have the same Casimirs for fixed parameters ~$m_1,...,m_{n-1},\mu$,
since such a multiplet can be obtained from  a main multiplet (of only one connected component)
depending on the parameters ~$m_1,...,m_{n-1},m_n$,
 by replacing the even number $2m_n$ with the odd number $\mu$.

There are ~$n$~ minimal UIRs for ~$sp(n,\bbr)$~ with conformal weights ~$d=\ha,1,...,\han$~ whose corresponding ERs
are denoted in the corresponding figures as ~$\chi^-_0,\chi^-_a,\ldots$. Note that
in each case the Knapp-Stein operator acting from the ER containing a minimal UIR  degenerates
to a differential operator of degree ~$n,n-1,\ldots,1$, (respectively to the above enumeration).
 Note further, that there is ~no~ differential operator with image that is an ER containing a minimal UIR - that can be used
as an equivalent definition of a minimal UIR. The only operator that acts ~to~ such an ER is the
conjugate non-degenerate Knapp-Stein integral operator.

Note that only the two singletons  with ~$d=\ha,1$~ are isolated points below the continuous unitary spectrum.

We should also note that the multiplets and \idos\ for
~$Sp(2r,\bbr)$~ are valid for  ~$Sp(r,r)$, though the representation
content is different \cite{Dobparel}.


 In the present paper we continued the programme outlined in
\cite{Dobinv} on the example of the non-compact group  $Sp(n,\bbr)$
(started already in \cite{Dobspn6}). Similar explicit descriptions
are planned for   other non-compact groups from which we have
considered so far the cases of $E_{7(-25)}$
\cite{Dobeseven},\footnote{For a different use of $E_{7(-25)}$, see,
e.g., \cite{CCM}.}\ $E_{6(-14)}$ \cite{Dobesix}, ~$SU(n,n)$
  \cite{Dobsunn}, ~$SO(p,q)$ \cite{Dobsopq} as
parabolically related  to ~$SO(n,2)$ \cite{Dobparel}. We plan also
to extend these considerations to the supersymmetric cases and also
to the quantum group setting  following as here the procedure of
\cite{Dob}  and its generalizations,   cf., e.g., \cite{Dobvar}.
Such considerations are expected to be very useful for applications
to string theory and integrable models, cf., e.g., \cite{Witten}.

\vfill

\fig{}{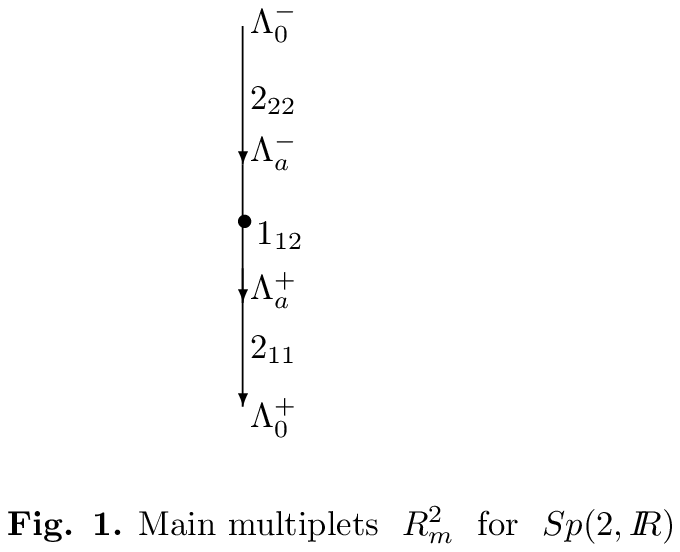}{80mm}
\fig{}{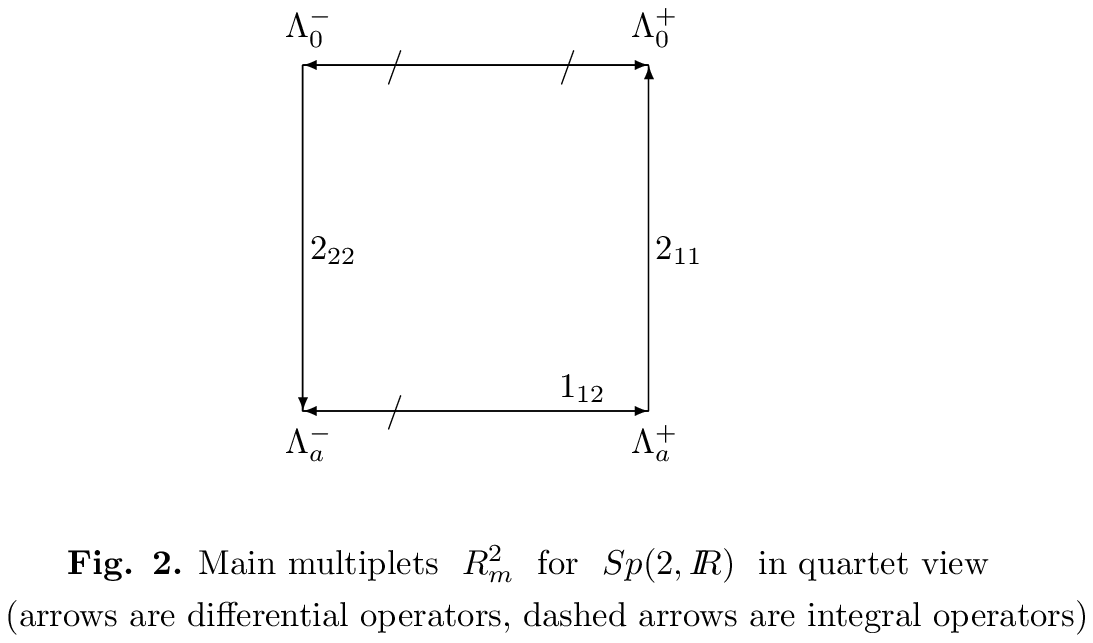}{90mm}
\fig{}{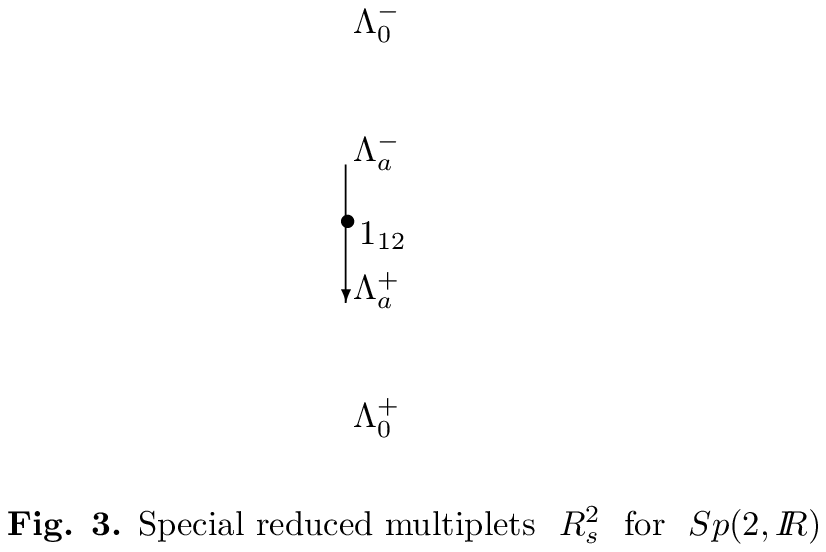}{80mm}

\fig{}{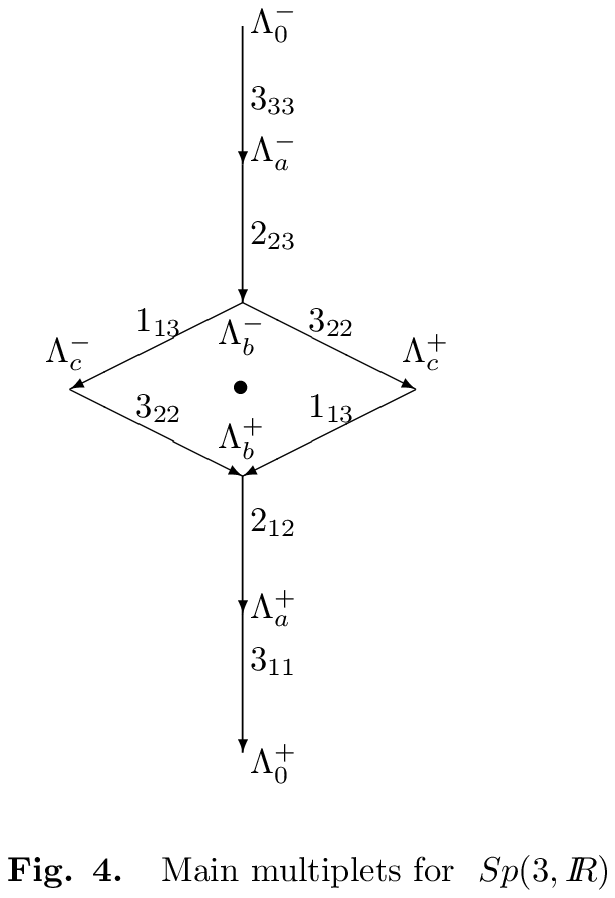}{90mm}
\fig{}{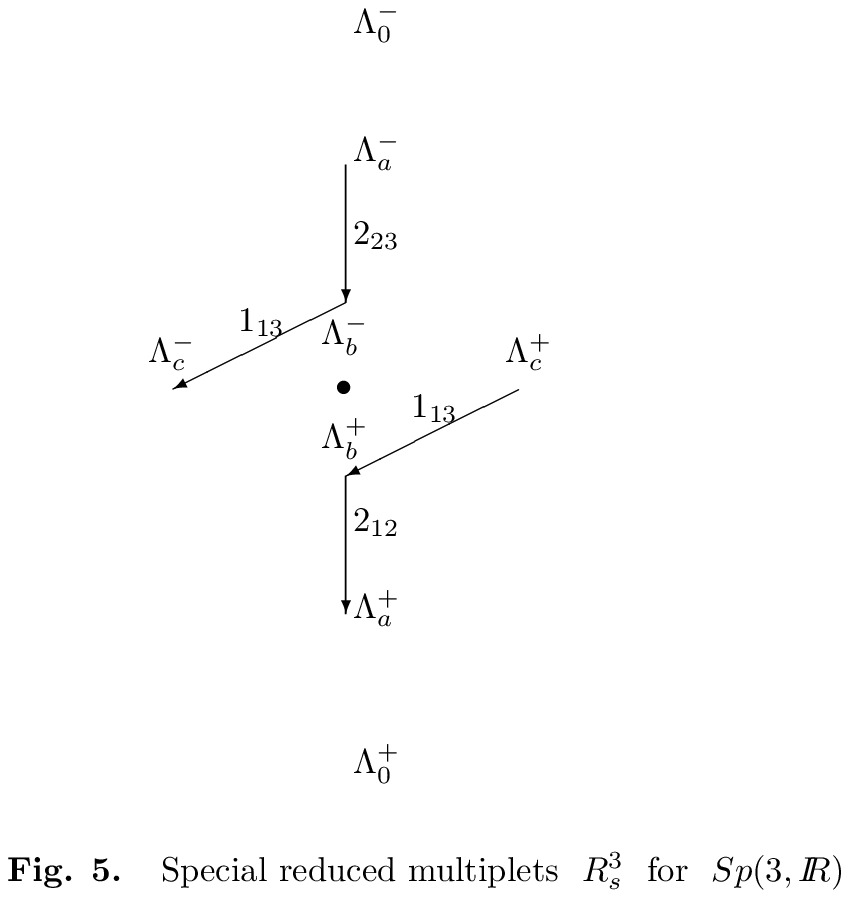}{90mm}
\fig{}{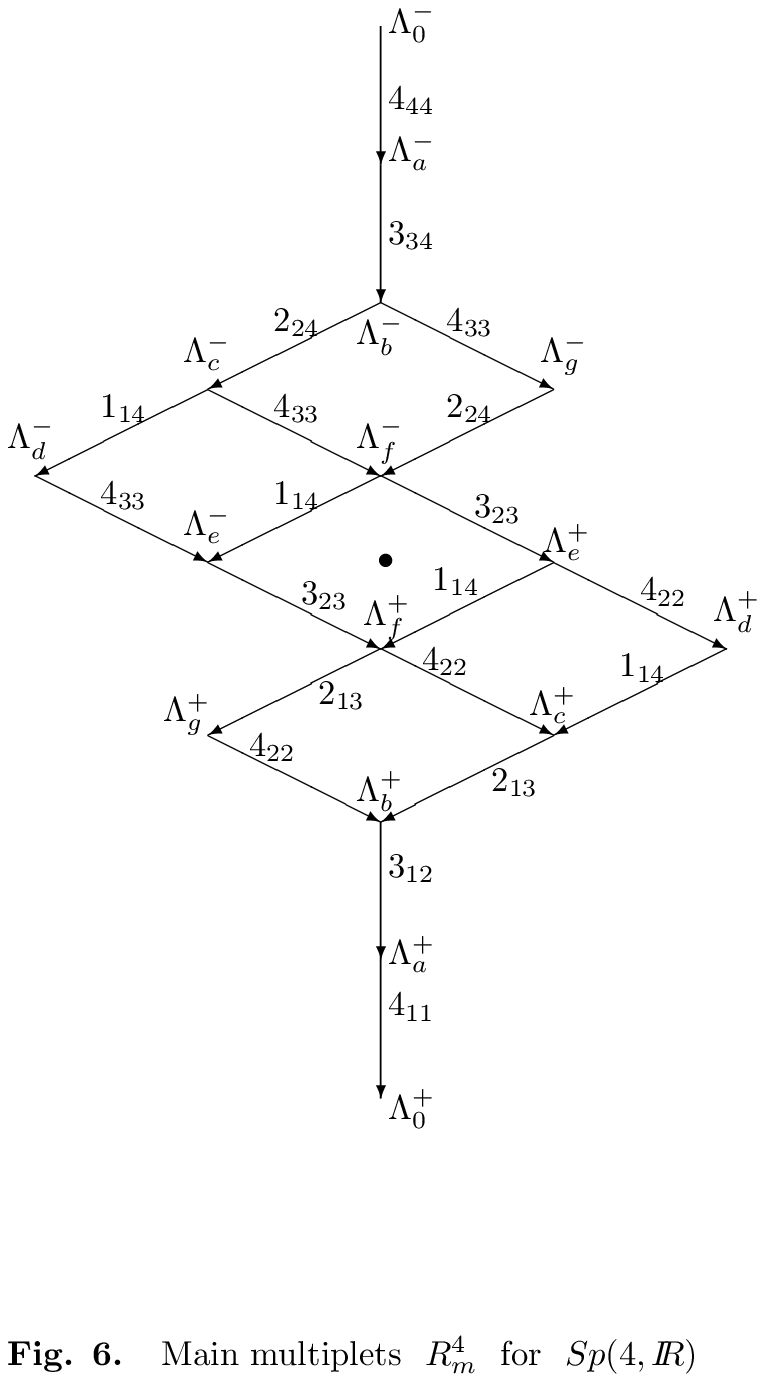}{90mm}
\fig{}{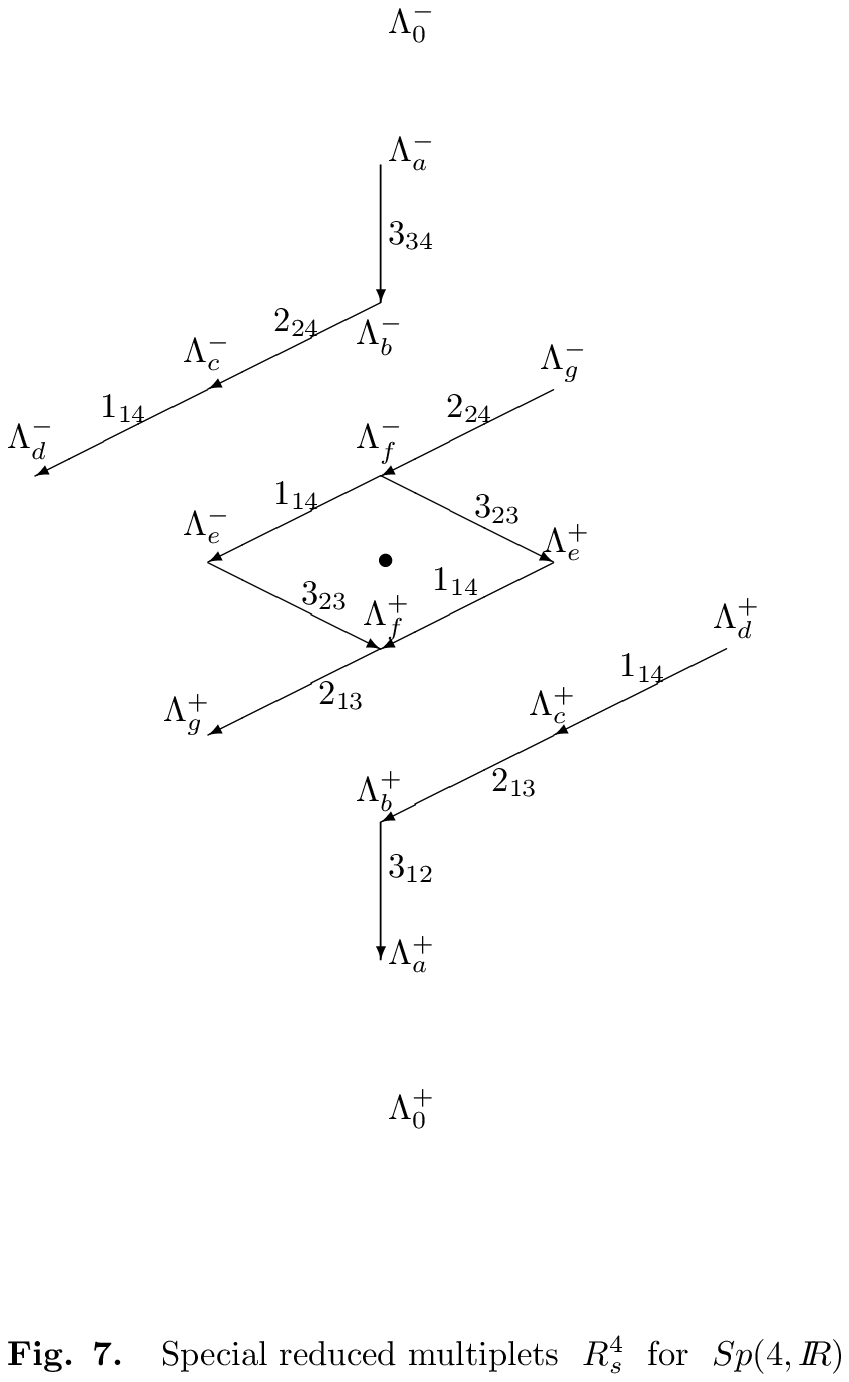}{80mm}

\fig{}{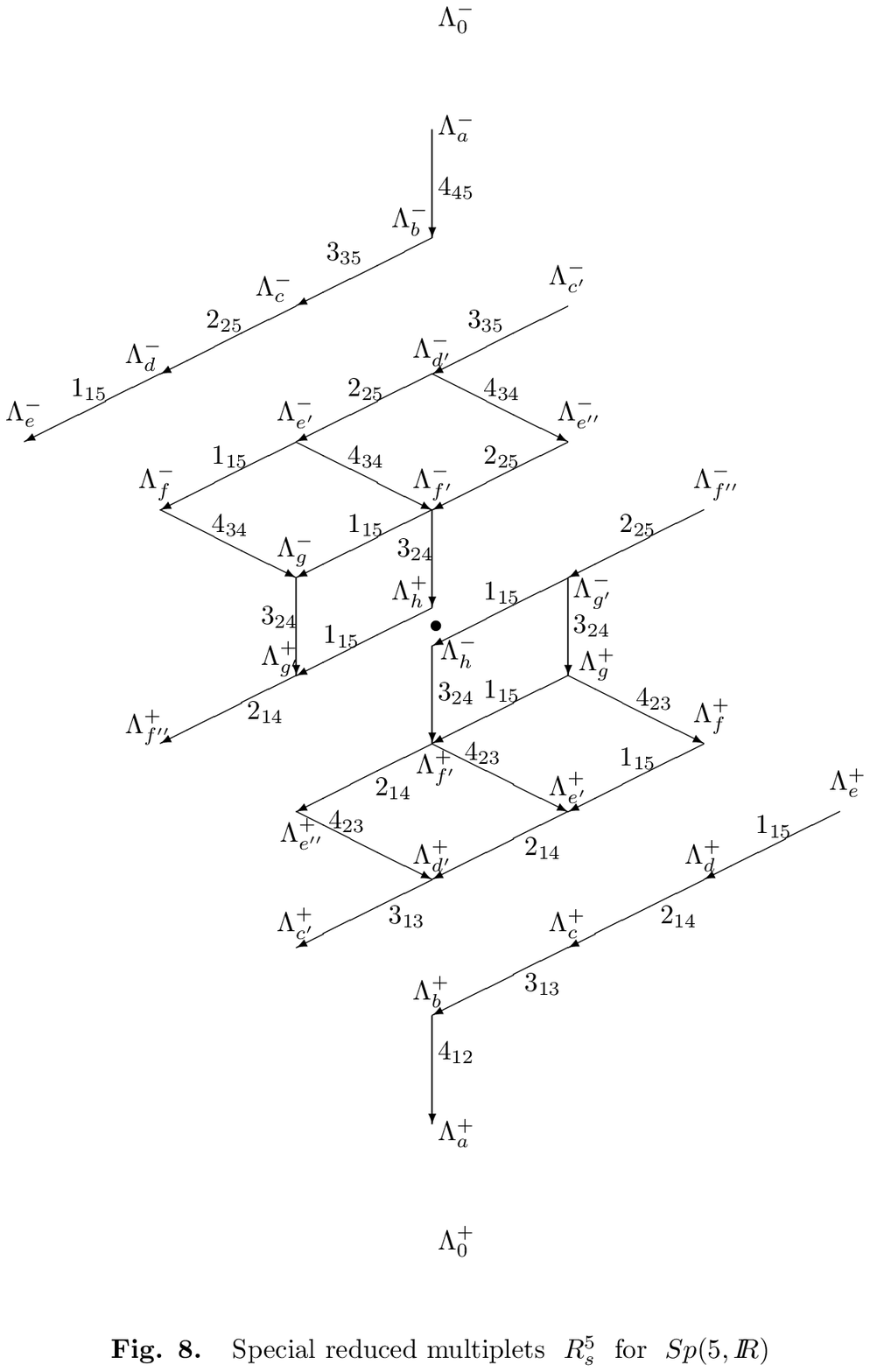}{120mm}
\fig{}{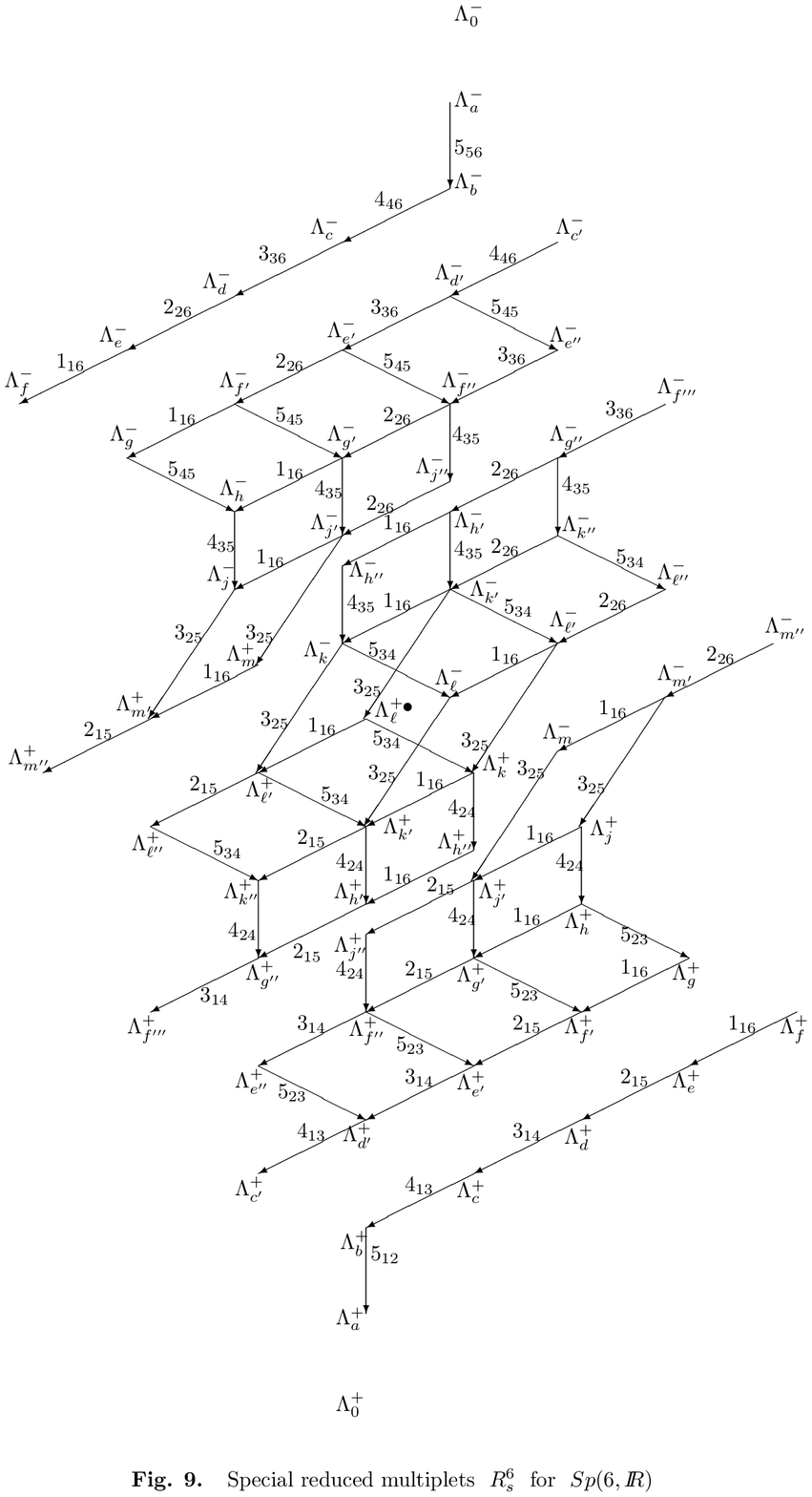}{120mm}


\begin{thebibliography}{B-B} 
\medskip
\begin{footnotesize}



\bibitem{Ter}J. Terning, {\it Modern Supersymmetry: Dynamics and
Duality}, International Series of Monographs on Physics \# 132,
(Oxford University Press, 2005).

\bibitem{Dobinv}V.K. Dobrev, 
Rev. Math. Phys. {\bf 20} (2008) 407-449; hep-th/0702152.

\bibitem{Dobeseven} V.K. Dobrev, 
J. Phys. {\bf A42} (2009) 285203,   arXiv:0812.2690
[hep-th].




\bibitem{FaKo} J. Faraut and A. Kor\'anyi,  {\it Analysis on Symmetric Cones}, Oxford
Mathematical Monographs, (Clarendon Press, Oxford, 1994).

\bibitem{Guna} M. Gunaydin, Mod. Phys. Lett. {\bf A8} (1993) 1407-1416.


\bibitem{Mackder} G. Mack and M. de Riese, J. Math. Phys. {\bf 48}
(2007) 052304;
hep-th/0410277v2.

\bibitem{Dobspn6}
V.K. Dobrev, Invariant Differential Operators for Non-Compact Lie
Groups: the Sp(n,R) Case, arXiv:1205.5521 [hep-th],
CERN-PH-TH/2012-143, to appear in    "Springer Proceedings in
Mathematics and Statistics" Vol. 36 (ISBN 978-4-431-54269-8).





\bibitem{HC}Harish-Chandra, "Representations of semisimple Lie groups:
IV,V", Am. J. Math. {\bf 77} (1955) 743-777, {\bf 78} (1956) 1-41.

\bibitem{DMPPT}V.K. Dobrev, G. Mack, V.B. Petkova, S.G. Petrova and
I.T. Todorov, {\it Harmonic Analysis on the $n$-Dimensional Lorentz
Group and Its Applications to Conformal Quantum Field Theory},
Lecture Notes in Physics, Vol. 63 (Springer, Berlin, 1977); V.K.
Dobrev, G. Mack, V.B. Petkova, S.G. Petrova and I.T. Todorov, Rept.
Math. Phys. {\bf 9}   (1976) 219-246; V.K. Dobrev and V.B. Petkova,
Rept. Math. Phys. {\bf 13}  (1978) 233-277.

\bibitem{Knapp}A.W. Knapp, {\it Representation Theory of Semisimple
Groups (An Overview Based on Examples)}, (Princeton Univ. Press,
1986).

\bibitem{Dob}V.K. Dobrev,
Rept. Math. Phys. {\bf 25}  (1988) 159-181; first as ICTP Trieste
preprint IC/86/393 (1986).

\bibitem{Dobmul}V.K. Dobrev, Lett. Math. Phys. {\bf 9}
(1985) 205-211; J. Math. Phys. {\bf 26} (1985) 235-251.

\bibitem{BGG}I.N. Bernstein, I.M. Gel'fand and S.I. Gel'fand,
Funkts. Anal. Prilozh. {\bf 5} (1) (1971) 1-9; English
translation: Funct. Anal. Appl. {\bf 5} (1971) 1-8.

\bibitem{Dix}J. Dixmier, {\it Enveloping Algebras}, (North Holland, New
York, 1977).

\bibitem{KnSt}A.W. Knapp and E.M. Stein,
Ann. Math. {\bf 93} (1971) 489-578; II : Inv.
Math. {\bf 60} (1980) 9-84.

\bibitem{Dobads} V.K. Dobrev,
J. Phys.  {\bf A39} (2006) 5995-6020; hep-th/0512354.

\bibitem{Dobpeds}V.K. Dobrev, 
J. Phys. {\bf A41} (2008) 425206; arXiv:0712.4375.

\bibitem{GeV}I.M. Gelfand, M.I. Graev and N.Y. Vilenkin, {\it
Generalised Functions}, vol. 5 (Academic Press, New York, 1966).

\bibitem{Koba}T. Kobayashi, Publ. RIMS, {\bf 47} (2011) 585-611.

\bibitem{Dirac}P.A.M. Dirac, J. Math. Phys. 4, 901 (1963).

\bibitem{Fro}C. Fronsdal, Rev. Mod. Phys. {\bf 37}, 221 (1965); Phys.
Rev. {\bf D10}, 589 (1974);   Phys. Rev. {\bf D12}, 3819 (1975).

\bibitem{Dobparel} V.K. Dobrev, Invariant Differential Operators for Non-Compact
Lie Algebras Parabolically Related to Conformal Lie Algebras, JHEP
to appear, arXiv:1208.0409 [hep-th], CERN-PH-TH/2012-215.

\bibitem{CCM} S.L. Cacciatori, Bianca L. Cerchiai, A. Marrani,
Magic Coset Decompositions, arXiv:1201.6314, CERN-PH-TH/2012-020.

\bibitem{Dobesix} V.K. Dobrev, 
Invited Lectures at 5th  Meeting  on Modern Mathematical Physics,
Belgrade, 6-17.07.2008, Proceedings, Eds. B. Dragovich, Z. Rakic,
(Institute of Physics, Belgrade, 2009) pp. 95-124; arXiv:0812.2655
[math-ph].


\bibitem{Dobsunn}V.K. Dobrev, Invariant Differential Operators for Non-Compact
Lie Groups: the Main SU(n,n)  Cases, Plenary talk at SYMPHYS XV,
(Dubna, 12-16.7.2011), to appear in the Proceedings.

\bibitem{Dobsopq} V.K. Dobrev, Conservation Laws for SO(p,q), arXiv:1210.8067
[math-ph hep-th math.RT],    Invited talk at XXIX International
Colloquium on  Group-Theoretical Methods in Physics, Chern Institute
of Mathematics, Nankai  Univ., China, August 20-26, 2012, to appear
in the Proceedings.



\bibitem{Dobvar}V.K. Dobrev, Phys. Lett. {\bf B186}, 43-51 (1987);  Phys. Part.
Nucl. {\bf 38}   (2007)  564-609, hep-th/0406154;~ V.K. Dobrev and
A.Ch. Ganchev, Mod. Phys. Lett. {\bf A3} (1988) 127-137; ~V.K.
Dobrev and P.J. Moylan, Phys. Lett. {\bf B315}   (1993) 292-298.

\bibitem{Witten}E. Witten, "Conformal Field Theory in Four and Six
Dimensions", arXiv:0712.0157.






\end{footnotesize}
\end{thebibliography}
\end{document}